\documentclass[onecolumn, 11pt]{article}

\usepackage{graphicx}
\usepackage[affil-it]{authblk}
\usepackage{fullpage}
\usepackage{setspace}
\usepackage{url}
\begin{document}

\title{The GAP Benchmark Suite}

\author{Scott Beamer, Krste Asanovi\'{c}, David Patterson\\sbeamer@eecs.berkeley.edu, krste@eecs.berkeley.edu, pattrsn@eecs.berkeley.edu}
\affil{Electrical Engineering \& Computer Sciences Department\\
    University of California\\
    Berkeley, California}

\maketitle

\begin{abstract}
We present a graph processing benchmark suite with the goal of helping to standardize graph processing evaluations. Fewer differences between graph processing evaluations will make it easier to compare different research efforts and quantify improvements. The benchmark not only specifies graph kernels, input graphs, and evaluation methodologies, but it also provides optimized reference implementations. These reference implementations are representative of state-of-the-art performance, and thus new contributions should outperform them to demonstrate an improvement.

The input graphs are sized appropriately for shared memory platforms, but any implementation on any platform that conforms to the benchmark's specifications could be compared. This benchmark suite can be used in a variety of settings. Graph framework developers can demonstrate the generality of their programming model by implementing all of the benchmark's kernels and delivering competitive performance on all of the benchmark's graphs. Algorithm designers can use the input graphs and the reference implementations to demonstrate their contribution. Hardware platform designers and performance analysts can use the suite as a workload representative of graph processing.
\end{abstract}

\section{Introduction}
Graph algorithms and their application are currently gaining renewed interest. Although the graph abstraction has been around centuries, it has become increasingly relevant, especially as social networks and their analysis grow in importance. Additionally, graph algorithms have found new applications in science and recognition. The result has been renewed interest in supporting graph processing well and for many sizes of graphs. Research is ongoing at all levels, including: applications, algorithms, implementations, frameworks, and even hardware platforms.

A lack of a standard for evaluations has clouded the results from the growing body of graph processing research. To compare a new result to prior work, ideally everything other than the new contribution should be the same. Unfortunately, there is often insufficient overlap between different published results to make such meaningful comparisons. Simple methodology differences (e.g., treating input edges as directed or undirected) can impact performance by more than the claimed improvement. Even for a particular well-known graph problem, there are often many variations (e.g., tracking parent vertices or vertex depths for breadth-first search) that alter what optimizations are possible. The input graphs themselves can be misleading, as similar or identical names can refer to significantly different graph datasets. A standard for graph processing evaluations could combat these problems.

Other graph processing evaluation mistakes would be harder to make if there is a well-known evaluation standard. With a standard set of diverse input graphs, if an optimization is only compatible with certain topologies, this weakness could be exposed. A standard set of input graphs could also prevent the use of graphs that are trivially small or unrealistically synthetically generated. A standard high-quality reference implementation could help discourage the use of low-performance baselines.

The Graph 500~\cite{Graph500} competition has been a great success for the graph community, so we use its strengths and expand upon them. Graph 500 has strong community adoption that has led to innovation in both algorithms and implementations. For example, the top finisher from the first competition in November 2010 is 5450$\times$ slower than the most recent top finisher (November 2016), and it would place no higher than $120^{th}$ in the current rankings. The biggest shortcomings of Graph 500 are its focus on one kernel (breadth-first search) and only one synthetic input graph topology (Kronecker). Although there are efforts to add additional kernels to Graph 500, they are still under review~\cite{Graph500-sssp}. The Problem-based benchmark suite~\cite{PBBS} is another noteworthy effort, but we intend to improve upon it by providing substantially higher performance reference code, using real-world graphs, and focussing on graph algorithms alone.

We present the GAP Benchmark Suite to ameliorate these evaluation concerns. The benchmark specifies graph kernels, input graphs, and measurement methodologies. The benchmark also includes an optimized reference implementation that is representative of state-of-the-art performance~\cite{gapbs}. To create our benchmark, we learned from the best practices of the community, and we co-developed this benchmark with our graph workload characterization~\cite{Beamer-locality}. An important lesson from our characterization is the importance of putting together a diverse workload consisting of multiple kernels and input graphs.


A key aspect of our benchmark suite is specifying the benchmark itself (this document). Other research efforts have released their code~\cite{COST,Galois,Ligra,GraphMat}, which eases comparisons to themselves, but the evaluator is still responsible for creating a workload and an evaluation methodology. Furthermore, since these frameworks were developed independently, they may require some modification to ensure the implementations are computing the same thing and using the same timing practices.

The benchmark specification and the reference implementation are two separate artifacts that can be used independently. By specifying the benchmark explicitly, any implementation on any platform that conforms to the benchmark's specifications can be compared. Consequently, these benchmark-compliant comparisons do not require the use of our reference implementation. Additionally, the reference implementation can be used to execute workloads other than the benchmark.

This benchmark suite can be used in a variety of settings. Graph framework developers can demonstrate the generality of their programming model by implementing all of the benchmark's kernels and delivering competitive performance on all of the benchmark's graphs. Algorithm designers can use the input graphs as a workload and the reference implementation as a baseline to demonstrate their contribution. Hardware platform designers and performance analysts can use the suite as a workload representative of graph processing.

Our reference implementation targets shared memory multiprocessor systems, but that platform is not required to use this benchmark. Our benchmark focuses on the execution of the graph kernels and does not consider the time required to load the graph data or to build the graph itself. Different platforms might load the graphs in different manners, but the input graphs, kernels, and methodologies are all platform-agnostic. Semi-external memory approaches could demonstrate their competitiveness with shared memory baselines. A distributed graph framework should prove itself worthy of cluster resources by substantially outperforming an optimized baseline running on a single node instead of only comparing against itself running on a single node~\cite{COST}.

\textit{Note: This benchmark incorporates feedback from the community and has undergone several revisions. Both this document as well as the reference implementations are versioned so future improvements will be clearly labelled.}

\section{Benchmark Specification}
\label{sec:spec}
This work is motivated by the evaluation shortcomings we observe in prior work. We created this benchmark based on our experiences competing in Graph500~\cite{Beamer-TR,Graph500}, performing our workload characterization~\cite{Beamer-locality}, and analyzing prior work~\cite{Beamer-thesis}. We designed this benchmark with the following goals in mind, each intended to solve existing evaluation problems:
\begin{itemize}
\item Explicit benchmark specifications to standardize evaluations in order to improve comparability and reduce confusion
\item Diverse representative workload to ensure evaluations have relevant target
\item High-quality reference implementation to ground evaluations with strong baseline performance
\end{itemize}


For clarity, we also include a brief summary of the notation we use. A graph $G(V,E)$ is made up by the set of its vertices $V$ and the set of its edges $E$. For conciseness, $n$ and $m$ are used to refer to the number of vertices ($|V|$) and the number of edges ($|E|$). The set of vertices with an edge from vertex $v$ (outgoing neighborhood) is $N^+(v)$ and the set of vertices with an edge to vertex $v$ (incoming neighborhood) is $N^-(v)$. If the graph is undirected, both neighborhoods are the same $N^+(v) = N^-(v)$.

We specify the benchmark by describing the graph kernels and input graphs. We also describe the required evaluation methodologies and provide rationale for our decisions.

\subsection{Graph Kernels}
We select six kernels based on how commonly they are used~\cite{Beamer-thesis}. These kernels are representative of many applications within social network analysis, engineering, and science. This set of kernels is computationally diverse, as it includes both traversal-centric and compute-centric kernels. Across the suite, different kernels will consider or ignore different properties of the graph including edge weights and edge directions. To remove ambiguity, we describe which variant of each graph problem we require and what constitutes a correct solution.

\begin{itemize}
\item \textbf{Breadth-First Search (BFS)} is a traversal order starting from a source vertex. BFS traverses all vertices at the current depth (distance from the source vertex) before moving onto the next depth. BFS is so fundamental, it is often implicit in other graph algorithms. We make BFS into a kernel by tracking the parent vertices during the traversal, akin to Graph 500~\cite{Graph500}. For any reached vertex, there is often more than one possible parent vertex, as any incoming neighbor with a depth one less than the reached vertex could be its parent. Multiple legal parent vertices cause there to be more than one correct solution to BFS from a given source vertex. For this reason, we define the correct solution to BFS starting from a \emph{source} vertex to be a \emph{parent} array that satisfies the following:
\begin{itemize}
  \item parent[source] = source
  \item parent[v] = -1 if v is unreachable from source
  \item if v is reachable and parent[v] = u, there exists an edge from u to v
  \item if v is reachable and parent[v] = u, depth[v] = depth[u] + 1
\end{itemize}
Some uses of BFS only track reachability or depth, but we choose to track the parent since it is the most helpful to the community. If we instead chose to track reachability, there would be no way to verify the traversal was performed in a breadth-first manner, because reachability only returns a boolean value for each vertex. Tracking depth would be a better choice than reachability since it allows for verifying the traversal, but BFS implementations that track depth can perform optimizations not possible in the more challenging applications of BFS that track parents~\cite{Duane}.

\item \textbf{Single-Source Shortest Paths (SSSP)} computes the distances of the shortest paths from a given source vertex to every other reachable vertex. The distance between two vertices is the minimum sum of edge weights along a path connecting the two vertices. We do not request the parent vertices, since BFS already provides this. There is a unique solution to our variant of SSSP, since the solution is the distances and not the shortest paths themselves. Although there may be more than one shortest path between two vertices, all shortest paths will have the same distance. All of our benchmark graphs have positive edge weights. We define the correct solution of SSSP to be the \textit{distance} array from a \textit{source} vertex such that:
\begin{itemize}
  \item distance[source] = 0
  \item distance[v] = $\infty$ (or some known sentinel value) if v is unreachable from source
  \item if v is reachable from source, there is a path of combined weight distance[v] from the source to v
  \item if v is reachable from source, there is no path of combined weight less than distance[v] from the source to v
\end{itemize}

\item \textbf{PageRank (PR)} computes the PageRank score for all vertices in the graph. The score ($PR$) for a vertex $v$ with a damping factor $d$ (0.85) is:
\[PR(v) = \frac{1-d}{|V|} + d\sum_{u\in N^-(v)}\frac{PR(u)}{|N^+(u)|}\]
For PR, we require the PageRank scores for all of the vertices such that a single additional iteration will change all of the scores by a sum of less than $10^{-4}$. More formally, if the benchmark kernel returns scores $PR_k$ and a classical implementation generates $PR_{k+1}$ in one iteration from $PR_k$, a solution is correct if it obeys the tolerance:
\[\sum_{v \in V}|PR_k(v) - PR_{k+1}(v)| < 10^{-4}\]
Selecting the required tolerance is a tradeoff between score convergence and execution time. We pragmatically select $10^{-4}$ since the scores will have mostly converged and it results in a reasonable number of iterations (5 - 20) for most graphs. Our tolerance bound also implicitly allows for a little numerical noise due to differences in accumulation order. We allow for more advanced implementations as long as they meet this bound and any changes to the graph or preprocessing optimizations are included in the trial time.

\item \textbf{Connected Components (CC)} labels all vertices by their connected component and each connected component is assigned its own unique label. If the graph is directed, we only require weakly connected components, so if two vertices are in the same connected component, it is equivalent to there being a path between the two vertices if the graph's edges are interpreted as undirected. Vertices of zero degree (disconnected) should each get their own label. To define correctness, we require the following equivalence relation:
\begin{itemize}
  \item vertices $u$ and $v$ have the same component label if and only if there exists an undirected path between $u$ and $v$
\end{itemize}

\item \textbf{Betweenness Centrality (BC)} approximates the betweenness centrality score for all vertices in the graph by only computing the shortest paths from a subset of the vertices. The betweenness centrality of a vertex $v$ is defined to be the fraction of shortest paths that pass through it. If $\sigma_{st}$ is the number of shortest paths between vertices $s$ and $t$ and $\sigma_{st}(v)$ is the number of those shortest paths that pass through $v$, the betweenness centrality score for $v$ is:
\[BC(v) = \sum_{s, t \in V, s \ne v \ne t}\frac{\sigma_{st}(v)}{\sigma_{st}}\]

For the approximation, the BC scores should be computed by considering the shortest paths from 4 different sources. These centrality scores should be normalized to one. We expect most implementations to utilize Brandes algorithm~\cite{Brandes} and perform 4 BFS traversals, but the use of Brandes algorithm is not required. The BC implementation should treat all input graphs as unweighted.

For correctness, we recommend comparing the output to the output from a simple correct implementation of an alternate algorithm using the same source vertices. In this case, the verifier may need to allow for a little numerical noise since the accumulations can happen in different orders.

\item \textbf{Triangle Counting (TC)} computes the total number of triangles in a graph. We define a \textit{triangle} to be three vertices that are directly connected to each other (clique of size 3). A triangle is invariant to permutation, so the same three vertices should be counted as only one triangle no matter the order in which they are listed. Additionally, for our definition of a triangle, we ignore the directions of the edges, so the input graph can be interpreted as undirected. For TC, the solution is unique, so comparing the result is trivial. Unfortunately, there is no easy way to verify the total number of triangles without actually computing it. For verification, we recommend comparing the result from the benchmark implementation with the result from an alternate implementation.
\end{itemize}

\subsection{Input Graphs} 
We select five input graphs for our benchmark and they are diverse in both topology and origin (synthetic versus real-world). Our real-world data models the connections between people, websites, and roads. The graph sizes are selected to be small enough to fit comfortably in most servers' memory yet large enough to be orders of magnitude bigger than the processors' caches. When selecting real-world benchmark graphs, we considered the ease of users acquiring the graph data and the graphs selected are amongst the easiest to obtain both in terms of licensing requirements and bandwidth availability.

\begin{itemize}
\item \textbf{Twitter} ($|V|$=61.6M, $|E|$=1,468.4M, directed) is an example of a social network topology~\cite{Twitter}. This particular crawl of Twitter has been commonly used by researchers and thus eases comparisons with prior work. By virtue of it coming from real-world data, it has interesting irregularities and the skew in its degree distribution can be a challenge for some implementations.
\item \textbf{Web} ($|V|$=50.6M, $|E|$=1,949.4M, directed) is a web-crawl of the .sk domain (\texttt{sk-2005})~\cite{UFL}. Despite its large size, it exhibits substantial locality due to its topology and high average degree.
\item \textbf{Road} ($|V|$=23.9M, $|E|$=58.3M, directed) is the distances of all of the roads in the USA~\cite{DIMACS}. Although it is substantially smaller than the rest of the graphs, it has a high diameter which can cause some synchronous implementations to have long runtimes.
\item \textbf{Kron} ($|V|$=134.2M, $|E|$=2,111.6M, undirected) uses the Kronecker synthetic graph generator~\cite{Kron} with the same parameters as Graph 500 (A=0.57, B=C=0.19, D=0.05)~\cite{Graph500}. It has been used frequently in research due to Graph 500, so it also provides continuity with prior work.
\item \textbf{Urand} ($|V|$=134.2M, $|E|$=2,147.4M, undirected) is synthetically generated by the Erd\H{o}s--R\'{e}yni model (Uniform Random)~\cite{Erdos}. With respect to locality, it represents the worst case as every vertex has equal probability of being a neighbor of every other vertex. When contrasted with the similarly sized kron graph, it demonstrates the impact of kron's scale-free property.
\end{itemize}

All of the graphs except road are unweighted, so weights must be added to the graphs before executing SSSP. To generate weights for the other graphs, we adopt the practice from Graph 500's SSSP proposal of using uniformly distributed integers from 1 through 255~\cite{Graph500-sssp}. We recommend using our reference code for generating these weights as it is deterministic.

\subsection{Measurement Methodologies}
Given the input graphs and kernels, we now specify the measurement methodologies. Building on the success of Graph 500~\cite{Graph500}, we reuse many of its best practices.

Executing only a subset of the benchmark kernels is allowed, as some users of this suite may only be investigating a single graph problem. However, it is highly recommended to always use all of the input graphs as they are selected to be diverse. A new innovation may not work well for all input graphs, so it is important to understand the topologies for which the innovation is advantageous.

Since our benchmark does not focus on graph loading or graph building, each trial can assume the graph is already loaded. For example, on a shared-memory multiprocessor, a loaded graph might reside completely in memory in the CSR format. There are no restrictions on the layout of the graph in memory, however, the same graph layout must be used for all kernels. Any optimizations done to the graph layout must not be beneficial for only one algorithm. It is legal to remove duplicate edges and self-loops from the graph. It is also legal to reorder the neighbors of a vertex. If the graph is anyhow transformed or converted from the format used for the other kernels, the graph conversion time must be included in the trial time. If the graph is relabelled, outputs of the graph kernel must use the original vertex labels and this label conversion time should also be included in the trial time.

We select the number of trials each kernel should run with the goal of minimizing evaluation time while capturing enough samples to return significant results. In general, the kernels can be grouped into two classes: ``single-source'' kernels take a source vertex to start from and ``whole-graph'' kernels process the entire graph every time in the same way. For single-source kernels (BFS, SSSP, and BC), there is naturally substantial variation in execution time so we execute 64 trials from different source vertices. The source vertices are randomly selected non-zero degree vertices from the graph, and we recommend the vertex selector from the reference code as it is deterministic. For the whole-graph kernels (PR, CC, and TC), we execute just enough trials to capture any performance non-determinism. We reduce the number of trials for TC since most implementations typically have little variance in execution time and the execution time for TC is typically orders of magnitude longer than the rest of the suite. The trial counts are summarized in Table~\ref{table:rules}.

\begin{table}
\begin{center}
\begin{tabular}{| l | l | l |}
\hline
Kernel & Trials & Output per Trial\\
\hline
\textbf{BFS} & 64 trials from 64 sources & $|V|$-sized array of 32-bit integers (vertex identifiers)\\
\textbf{SSSP} & 64 trials from 64 sources & $|V|$-sized array of 32-bit integers (distances)\\
\textbf{PR} & 16 trials & $|V|$-sized array of 32-bit floating point numbers\\
\textbf{CC} & 16 trials & $|V|$-sized array of 32-bit integers (component labels)\\
\textbf{BC} & 16 trials each from 4 sources & $|V|$-sized array of 32-bit floating point numbers\\
\textbf{TC} & 3 trials & 64-bit integer (number of triangles)\\
\hline
\end{tabular}
\caption{Trial counts and output formats for benchmark kernels}
\label{table:rules}
\end{center}
\end{table}

Each trial of a kernel should be timed individually and it should include every aspect of its execution. Each trial can assume the graph is already loaded. Any time to construct data structures other than the graph used by the kernel, including memory allocated for the solution, must be included in the trial time. Additionally, the graph is the only data structure that can be reused between trials, as the purpose of repeated trials is to measure variance, not to amortize optimizations.

For each kernel, there must be only one implementation used for all input graphs. If different approaches will be better for different graph topologies, they should be combined into a hybrid implementation that includes a runtime heuristic (included in kernel time) to decide which approach to use. The same restriction applies to tuning parameters. None of the kernels may take parameters specific to the input graph with the exception of a $\Delta$ parameter for SSSP. We allow $\Delta$ because it is difficult to achieve high-performance on SSSP without it~\cite{delta-kamesh}. Fortunately, when SSSP is used in practice, $\Delta$ is available since its determining factors (graph diameter and edge weight distribution) are known within an application domain. A kernel can take tuning parameters specific to the hardware platform that are the same for all input graphs.

The particular output formats are summarized in Table~\ref{table:rules}. Our benchmark allows for the use of 32-bit vertex identifiers, which is in contrast to Graph 500 which requires at least 48-bit vertex identifiers. Unlike Graph500, for our benchmark, the graphs are of a known size and 32 bits comfortably accommodates them. Using larger identifiers unnecessarily penalizes the performance of cache-based systems, as none of the graphs will utilize more than 28 bits per identifier. However, benchmark implementations must support graphs with more than $2^{32}$ edges, and this is typically accomplished with 64-bit pointers. This mix of 32-bit and 64-bit types may seem inconsistent, but this is common practice due to the size of graphs that can fit within the memory of today's multiprocessor platforms. When the memory capacity of systems increases substantially, the input graphs will need to be expanded and this requirement may be increased.


\section{Reference Code}
In addition to the benchmark specification, we also provide a reference implementation that includes all six benchmark kernels. This implementation serves multiple purposes. At a minimum, the implementation is compliant with the benchmark specifications and can serve as a high-performance baseline for evaluations. The implementation implements state-of-the-art algorithms that provide it with competitive performance that is the best for some of the kernels. Additionally, the implementation serves an educational purpose, as it clearly demonstrates how to implement some leading algorithms that may not be clearly described by their original algorithmic descriptions. Furthermore, the implementation can help researchers, as much of the infrastructure can be reused as a starting point for high-performance graph algorithm implementations.

To serve all of these purposes, extra effort has been taken to improve code quality. Doing so not only helps those that read it, but it also makes it easier to modify and more portable. Our code follows Google's C++ style guide~\cite{google-style} and uses many best practices for C++ software engineering~\cite{GangOf4,effective-cpp}. Our code leverages many of the features of C++11 that allow us to program safely without any loss of performance. For example, the kernel implementations and most of the core infrastructure do not perform manual memory management or even use pointers. For parallelism, we leverage OpenMP (version 2.5 or later) and restrict ourselves to its simpler features in order to keep our code portable. We have successfully built and run our code on the x86, ARM, SPARC, and RISC-V ISA's using the gcc, clang, icc, and suncc compilers.

The core infrastructure for our reference code includes support for loading graphs from files, synthetically generating graphs, and building graphs in memory. We support multiple file formats, including the popular METIS~\cite{METIS} and Matrix Market formats~\cite{MTX}. As a fallback method for importing graphs into our infrastructure, we also support an extremely simple plain text format that is easy to convert to. Once our infrastructure has built a graph, it can serialize it to a file so later invocations can simply load the serialized graph directly into memory to skip building the graph. Loading the serialized graph saves time and reduces peak memory consumption. We also provide synthetic graph generators for urand and kron. The synthetic graph generators take advantage of seeding and C++11's strict random number generator specifications to deterministically produce the same graph in parallel even on different platforms or with different numbers of threads.

Our reference code includes testing throughout. In addition to testing the code for loading a graph from a file, generating a graph synthetically, or building a graph in memory, our kernel implementations can verify the correctness of their outputs. When there is more than one correct output (BFS, PR, and CC), our verifiers test the output for the properties of a correct solution. For the other kernels (SSSP, BC, and TC), we compare the output from the benchmark implementation to the output of a simple serial implementation that implements the kernel with a different algorithm.

Our reference code also provides scripts to automate the process of executing the benchmark suite. These scripts can build the benchmark's graphs which includes downloading the needed real-world data (twitter, web, and road). To execute the benchmark suite, the scripts run the reference implementations with the correct parameters for a benchmark-compliant execution. Furthermore, these scripts instructively demonstrate how to collect the input graphs and execute the kernels.

Here we describe some of the most noteworthy or novel optimizations employed by our kernel implementations, but we of course recommend examining the code itself~\cite{gapbs} to answer detailed questions.

\begin{itemize}
\item \textbf{Breadth-First Search (BFS)} implements the current state-of-the-art direction-optimizing algorithm~\cite{Beamer:SC-2012}. We perform an additional optimization to reduce the amount of time spent calculating how many edges exit the current frontier. Calculating the number of edges in the frontier is done during the top-down steps to determine if the frontier is large enough to justify switching to the bottom-up approach. Computing this total can have many irregular memory accesses, as it is summing the degrees of an unordered list of vertices. Our optimization is to add a step before the search that stores the degree of each vertex in the parent array as a negative number and this takes little time as the vertices can be done in order (great spatial locality). The previous convention of -1 representing an unvisited vertex is now revised to be any negative number represents an unvisited vertex. With this new encoding, during the top-down steps, the degree of a newly reached vertex is already known because the parent array was just checked to see if that vertex was unvisited. This optimization cuts out totaling the degrees of the frontier because it can now be done during the search. For low-diameter graphs, this yields a modest performance improvement, but for high-diameter graphs that will only use the top-down approach, this can yield a speedup of nearly 2$\times$. Our implementation does a final pass at the end to hide this optimization, and it sets any negative parent array value to -1.

To verify the output of our BFS implementation, we test for the properties this benchmark specifies for BFS. We check that the parent of the source is the source. We check that there is an edge from the parent of $v$ to $v$. Finally, we use a trivial serial BFS implementation to obtain the depths of all vertices from the source, and we use those depths to check that parents have depth one less than their children.

\item \textbf{Single-Source Shortest Paths (SSSP)} implements the $\Delta$-stepping algorithm~\cite{Delta} with some implementation optimizations from Madduri et al.~\cite{Madduri2007}. A common challenge for implementing delta-stepping is implementing the bins used to radix sort the vertices by distance. These bins are challenging to implement correctly because they need to be high-performance and support concurrent insertions. This task is further complicated by determining the sizes of these shared bins. If the bins are allowed to grow, they must be able to grow concurrently. If the bins are sized sufficiently large to not need to grow, there is the possibility of substantial wasted memory capacity.

We sidestep the challenge of implementing high-performance, concurrent, resizable bins by using thread local bins. Since the bins are thread local, there are no atomicity concerns and we can use existing serial resizable containers. To allow for work-sharing across threads, we use a single shared bin. This shared bin holds the vertices within the current minimum distance range. The shared bin is easy to implement since the threads only need read-only access. Every iteration, we copy all of contents of the thread-local bins for the minimum distance range into the shared bin. Since these aggregations are done in bulk, there is much less contention, which improves performance. Since there is only a single shared bin, it can be over-allocated such that it does not need to grow. Our design decision to use thread-local bins greatly simplifies our implementation as there are no longer parameters to tune for bin size or the number of bins to pre-allocate.

To verify the output of our SSSP implementation, we compare the distances to the output of a simple serial implementation of Dijkstra's algorithm. If the edge weights and distances are floating point, there are fortunately no concerns about numerical reproducibility. All known practical SSSP implementations add the edge weights in the same order, starting from the source vertex continuing along the shortest path.

\item \textbf{PageRank (PR)} uses the naive iterative approach that is quite similar to sparse matrix vector multiplication (SpMV). To avoid the use of atomic memory operations, we perform all updates in the pull direction. Unlike the rest of our implementations in this suite, for PR we deliberately choose to not implement the most sophisticated state-of-the-art algorithm in order to be easily comparable. PR is the most commonly used benchmark, and most often it is implemented in this same classical way. By using the classic approach, our implementation can be directly compared to many implementations since they will both perform the same amount of algorithmic work. Furthermore, many optimized implementations not only change the amount of algorithmic work, but in actuality do not obtain the tolerance bounds they advertise. Our benchmark specification allows for optimized PageRank implementations, but they must meet the tolerance bounds. Our implementation computes the tolerance every iteration and can be sure it is met when it decides to terminate.

We verify the output of our PR implementation by performing an additional iteration and computing the sum of the scores' changes. To contrast from our reference implementation, our verifier's implementation is serial and performs updates in the push direction.

\item \textbf{Connected Components (CC)} implements the Shiloach-Vishkin~\cite{SV-Alg} algorithm with parallelization techniques from Bader et al.~\cite{Bader2005}.

We verify the result of our CC implementation by checking for the equivalence stated in the CC specification. We accomplish this by performing a single traversal for each label. During each traversal, we check that a different label is not encountered. After all of these traversals, we assert that every vertex has been reached by a traversal. If two components share the same label, they will have unreached vertices because there is only a single traversal per label. Vertices of zero degree have their own label so they will each be traversed trivially.

\item \textbf{Betweenness Centrality (BC)} implements the Brandes~\cite{Brandes} algorithm with the lock-free improvements from Madduri et al.~\cite{Madduri2009}. To obtain a slight speedup and a large memory savings, we record the successors identified during the BFS pass in a bitmap rather than many lists whose total capacity is $O(|E|)$.

We verify the output of our BC implementation by comparing it to the output from a simple serial implementation. Our verifier implementation also uses Brandes algorithm, however, it is implemented in a different way and it does not record the successors during the BFS pass.

\item \textbf{Triangle Counting (TC)} implements two well-known optimizations~\cite{Chu-triangle}. To count triangles, we sum the sizes of the overlap between a vertex's neighbor list and its neighbors' neighbor lists. Our first optimization leverages our neighbor lists being sorted and terminates these intersection computations once the triangles found will not obey the invariant of $u > v > w$. Terminating these intersection computations early prevents each triangle from being counted six times. Our second optimization relabels the graph by degree, so when the first ordering optimization is applied, we get additional algorithmic savings. Relabelling the graph is compute-intensive, but counting triangles exactly is also compute-intensive so the relabelling optimization is often worthwhile even for a single execution. We find relabelling the graph is typically beneficial for scale-free graphs, and we use a heuristic to decide when to do it. Our heuristic samples the degrees of vertices and decides if the degree distribution is sufficiently skewed by comparing the sample's median degree with the graph's average degree.

We verify the total from our TC implementation by comparing it the total returned by a trivial serial implementation that uses a standard library implementation of set intersection. Our simple verifier implementation counts each triangle six times so we divide its initial total by six.
\end{itemize}

\appendix
\section{Change Log}
\begin{itemize}
\item \textbf{Changes since v3}
\begin{itemize}
  \item Introduction: clarified possible use cases for suite
  \item Benchmark Specification: clarified requirements for correct solutions
  \item Reference Code: move to version v1.0 in GitHub repo
  \item Reference Code: includes automation for executing benchmark
  \item Reference Code: support for more input file formats
  \item Wording improved throughout with content from dissertation 
\end{itemize}

\item \textbf{Changes since v2}
\begin{itemize}
  \item Re-uploaded, build didn't include spec.tex
\end{itemize}

\item \textbf{Changes since v1}
\begin{itemize}
  \item Benchmark Specification: converted trials and output into table format
  \item Reference Code: description of testing and verifier implementations
  \item Reference Code: move to version v0.7 in GitHub repo
  \item Acknowledgements: added section
  \item General wording and grammar fixes throughout
\end{itemize}
\end{itemize}

\section{Acknowledgements}
We thank Jinho Lee of Seoul National University for reporting a bug in BC for directed graphs.

We thank Marco Minutoli of PNNL for contributing support for importing graphs in the METIS format.

We thank Arnau Prat of Universitat Polit\`{e}cnica de Catalunya for contributing a bug report and a fix.

We thank Kirk Dunkelberger of Northrop Grumman for reporting two bugs.

Research partially funded by DARPA Award Number HR0011-12-2-0016, the Center for Future Architecture Research, a member of STARnet, a Semiconductor Research Corporation program sponsored by MARCO and DARPA, and ASPIRE Lab industrial sponsors and affiliates Intel, Google, Huawei, Nokia, NVIDIA, Oracle, and Samsung. Any opinions, findings, conclusions, or recommendations in this paper are solely those of the authors and does not necessarily reflect the position or the policy of the sponsors.

\bibliography{references}

\end{document}